\documentclass[10pt,twocolumn,letterpaper]{article}

\usepackage{iccv}
\usepackage{times}
\usepackage{epsfig}
\usepackage{graphicx}
\usepackage{amsmath}
\usepackage{amssymb}
\usepackage{amssymb}% http://ctan.org/pkg/amssymb
\usepackage{pifont}% http://ctan.org/pkg/pifont
\usepackage{makecell}

\newcommand{\cmark}{\ding{51}} %
\newcommand{\xmark}{\ding{55}} %

\usepackage[breaklinks=true,bookmarks=false]{hyperref}

\iccvfinalcopy % *** Uncomment this line for the final submission

\def\iccvPaperID{****} % *** Enter the ICCV Paper ID here

\ificcvfinal\fi

\begin{document}

%%%%%%%%% TITLE
\title{The First Vision For Vitals (V4V) Challenge for Non-Contact Video-Based Physiological Estimation}

\author{Ambareesh Revanur\\
Robotics Institute\\
Carnegie Mellon University\\
\and
Zhihua Li\\
Dept. of Computer Science\\
Binghamton University\\

\and
Umur A. Ciftci\\
Dept. of Computer Science\\
Binghamton University\\

\and
Lijun Yin\\
Dept. of Computer Science\\
Binghamton University\\

\and
L{\'a}szl{\'o} A. Jeni\\
Robotics Institute\\
Carnegie Mellon University\\

}

% \vspace{-300pt}
\maketitle
\ificcvfinal\thispagestyle{empty}\fi

%%%%%%%%% ABSTRACT
\begin{abstract}

Telehealth has the potential to offset the high demand for help during public health emergencies, such as the COVID-19 pandemic. Remote Photoplethysmography (rPPG) - the problem of non-invasively estimating blood volume variations in the microvascular tissue from video - would be well suited for these situations. Over the past few years a number of research groups have made rapid advances in remote PPG methods for estimating heart rate from digital video and obtained impressive results. How these various methods compare in naturalistic conditions, where spontaneous behavior, facial expressions, and  illumination changes are present, is relatively unknown. To enable comparisons among alternative methods, the 1st Vision for Vitals Challenge (V4V) presented a novel dataset containing high-resolution videos time-locked with varied  physiological signals from a diverse population. In this paper, we outline the evaluation protocol, the data used, and the results. V4V is to be held in conjunction with the 2021 International Conference on Computer Vision \footnote{\url{https://vision4vitals.github.io}}. %This topic is germane to both computer vision and multimedia communities. For computer vision, it is an exciting approach to longstanding limitations of vital signs estimating approaches. For multimedia, remote vital signs estimation would enable more powerful applications.

\vspace{-3mm}
\end{abstract}

%%%%%%%%% BODY TEXT
\section{Introduction}
There has been a tremendous interest in the extraction of human physiological signals using just facial videos. Computer vision based physiological extraction has been gaining momentum steadily because this technology has significant benefits over traditional contact-based measurements. Firstly, these methods allow reliable estimation of Heart Rate (HR) and Respiration Rate (RR) in absence of specialized equipment such as electrocardiogram (ECG). These methods depend only on the video feed recorded from a general RGB camera readily available in a commodity smartphone. Secondly, these methods operate without any contact with the subject. Hence, video-based physiology estimation promotes social distancing and is more patient-friendly than contact-based devices. Thirdly, these methods aid in remote diagnosis of patients located in remote areas where quality healthcare facilities are limited. These methods have a wide range of applications including telehealth, deep fake detection, affective computing, human behavior understanding, and sports. 

\begin{figure}[t]
\includegraphics[width=\columnwidth]{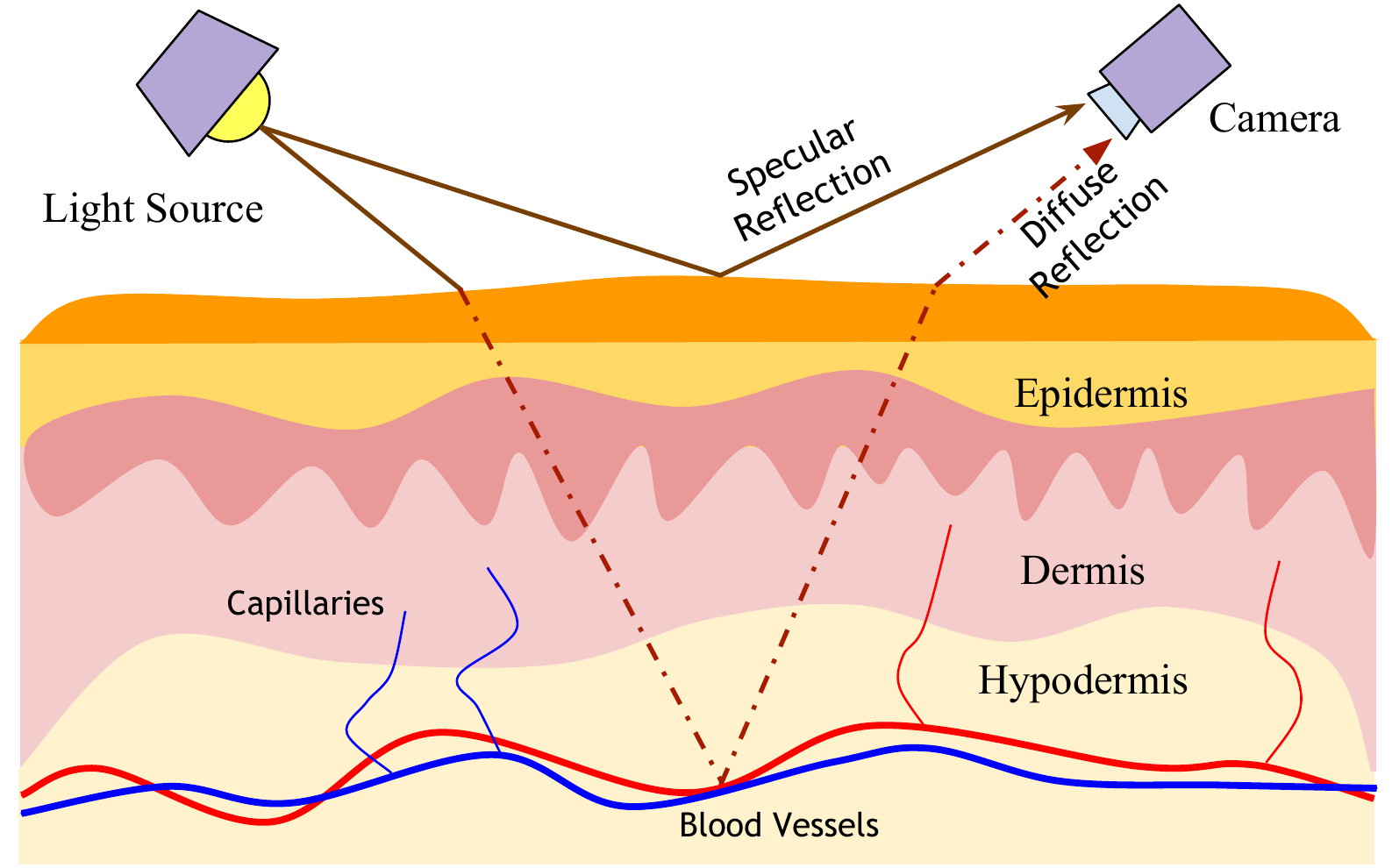}
\caption{Skin reflection model \cite{wang2016algorithmic}}
\label{fig:sdrm}
\vspace{-4mm}
\end{figure}

\begin{table*}[t]
    \centering
    \setlength\tabcolsep{2.5pt}
     \caption{Summary of previous challenges and recent methods in the area of video-based non-contact physiology estimation. Many of the previous evaluation protocols use non-overlapping segments. In V4V 2021, we use a continuous frame-level error measurement metrics.}
    \begin{tabular}{|c|c|c|c|c|}
    
    \hline
         & HR & RR & Datasets used & Evaluation Protocol   \\
         \hline
         \multicolumn{5}{|c|}{\textbf{Related Challenges}} \\
         \hline
         \makecell{The 1st Remote Physiological Signal Sensing \\ (CVPR'20) \cite{repss2020} }  & \cmark & \xmark & VIPL-HR-v2, OBF  & 10s non-overlapping segment \\
         \hline
         \makecell{The 2nd Remote Physiological Signal Sensing \\ (ICCV'21)}  & \cmark & \cmark & VIPL-HR-v2, OBF  & \makecell{Continuous metric: \\ Inter-beat interval} \\
         \hline
         \multicolumn{5}{|c|}{\textbf{Recent Methods}} \\
         \hline
         MTTS-CAN (NeurIPS'20) \cite{neurips2020multi} & \cmark & \cmark & AFRL, MMSE-HR & 30s non-overlapping segment \\
         \hline
         Feature Disentanglement (ECCV'20) \cite{eccv_20_video} & \cmark & \cmark & OBF, VIPL-HR, MMSE-HR & 30s non-overlapping segment \\
         \hline
         \multicolumn{5}{|c|}{\textbf{1st Vision-For-Vitals Challenge}} \\
         \hline
         Ours (ICCV'21) & \cmark & \cmark & \textit{V4V dataset} & \makecell{Continuous metric: \\ Frame-level HR/RR}  \\
         \hline
         
    \end{tabular}
   
    \label{tab:related}
\end{table*}

% \vspace{2mm}

Clinicians use FDA-approved devices such as an electrocardiogram (ECG), a chest belt, and a photoplethysmography (PPG) device for extracting human physiology signals such as HR and RR. Since PPG is closely related to the V4V challenge, we describe the functioning of a PPG device. It is a contact-based device capable of extracting the subtle imperceptible color changes induced as a result of periodic changes in the volume of blood flowing in the underlying skin tissues. In a simplified setup, it consists of a light emitter and receiver. While the emitter is used to focus the light beam on the skin tissue, the receiver is used to record the intensity of light transmitted back to the PPG device. It is known that the absorption spectrum of (oxy-) hemoglobin lies in the color band corresponding to green \cite{green2008remote}. Accordingly, the emitter and receiver are designed to capture the periodic color variations in the frequency range of heart rate. Studies have also shown that the respiration rate can be extracted either through motion analysis \cite{janssen2015video} or using the PPG signals \cite{tarassenko2014non3in1}. %Extracting RR is more challenging than the extraction of HR because the frequency range of RR is very low. 
However, a PPG device is a contact-based method and does not offer the attractive benefits offered by a non-contact-based method. To this end, several video-based non-contact remote physiology estimation methods have been advanced.

% \vspace{2mm}

The video-based physiology estimation methods \cite{ica2010non,green2008remote,chrom2013robust,wang2016algorithmic} exploit the reflectance properties of the skin (typically facial region) with an aim to extract the human physiological signals. Often, the skin tissue is modeled under Shafer’s Dichromatic Reflection Model (DRM) \cite{wang2016algorithmic} that provides a way to model the behavior of the light energy incident on surfaces. As shown in Fig. \ref{fig:sdrm}, the light incident on the skin tissue reflects back to the camera as two components - specular and diffuse reflectance. A fraction of the incident light energy that is reflected right off the skin surface is the specular component. This appears as a glossy/shiny reflection on the image captured using the camera. The diffuse reflectance is the light component that passes through the blood-rich tissues under the skin and is then transmitted out. Therefore, the diffuse component contains the signature of physiological signals, while the specular component does not. Similar to PPG, a careful analysis of the variations in the diffuse component of the reflected light shows a pulsatile signal in the frequency range of heart rate. Therefore, video-based physiological measurements techniques are often referred to as Remote-PPG (rPPG) methods.

% \vspace{2mm}

Several methods have been advanced for the extraction of rPPG signals \cite{chrom2013robust,wang2016algorithmic,ica2010non,green2008remote,eccv2018deepphys,neurips2020multi,eccv_20_video,eccv2020meta,cvpr2021dual}. Owing to the advancements in computer vision and deep learning, remarkable results have been achieved on the task of human physiology estimation. However, there are two drawbacks with existing methods. First, it is not clear how these various methods compare in naturalistic conditions, where spontaneous movements, facial expressions, and illumination changes are present. Second, most previous benchmarking efforts focused on posed situations. No commonly accepted evaluation protocol exists for estimating vital signs in spontaneous behavior with which to compare them. Therefore, in the 1st Vision-For-Vitals 2021 challenge, we introduce a new dataset called \emph{Vision-For-Vitals (V4V) dataset} which includes challenging elements such as spontaneous behavior. We also contribute a new evaluation metric for a stronger benchmarking on the \textit{V4V dataset}.

\section{Related works}

Table \ref{tab:related} summarizes the related challenges and most recent methods in the area of remote video-based human physiological estimation.

\vspace{1mm}

\noindent \textbf{Related challenges.} In conjunction with CVPR'20, the 1st Remote Physiological Signal Sensing (RePSS) challenge was organized for estimation of heart rate using RGB videos. The challenge consisted of a training dataset that was drawn from VIPL-HR-v2 \cite{niu2018vipldata} and a test set that was drawn from VIPL-HR-v2 \cite{niu2018vipldata} and the OBF \cite{obfdataset} datasets. There were about 2500 samples, each 10 seconds long, in the training dataset. In order to test the efficacy of different methods employed in the challenge, the organizers used a segment-level evaluation protocol, i.e., one heart rate prediction for a 10s segment.

The 2nd Remote Physiological Signal Sensing (RePSS) challenge was held recently\footnote{in conjunction with ICCV'21}, and it included samples drawn from VIPL-HR-V2 and OBF. In this challenge, the organizers introduced an inter-beat-interval (IBI) based metric for measuring the performance of the participants' methods. Unlike the segment-level metric used in the 1st RePSS, the IBI-based metric is more granular as it penalizes missed/extra heartbeat predictions. In the 2nd RePSS challenge, the organizers also included respiration rate as a separate challenge track. Our V4V challenge is similar in spirit to the 2nd RePSS as we used granular evaluation metrics and have both tracks (HR and RR). Further, we introduce a newly curated dataset that contains challenging elements such as spontaneous behavior and varied physiological signals as part of our V4V challenge.

\begin{figure}[t]
    \centering
    \includegraphics[width=\columnwidth]{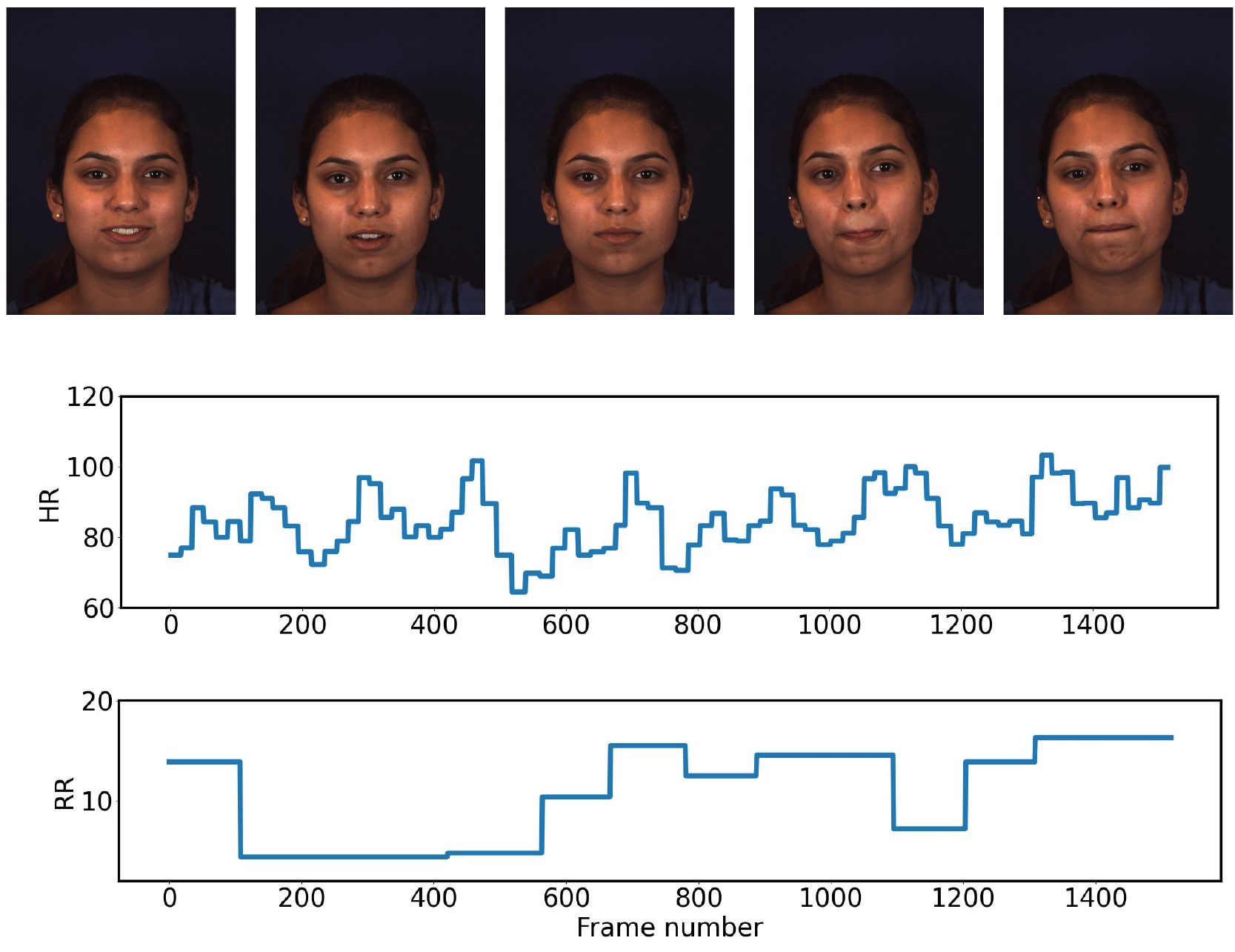}
    \caption{Subject participating in a cold presser (T8) task}
    \label{fig:emotion_f74}
\end{figure}

\vspace{1mm}

\noindent \textbf{Methods.} Traditional rPPG extraction methods typically involve two stages. The signal is first extracted based on the rPPG principles, and then the signal processing methods are used to compute the HR and RR. In \cite{green2008remote}, the pixels in the green channel are used to extract the physiological signals since it contains a strong signature of the pulsating rPPG signal. Some methods such as \cite{bkf,ica2010non, npjBiases2021} use an ICA method to determine the underlying rPPG signal followed by 3rd order Butterworth bandpass filtering to obtain the power spectrogram whose peak corresponds to the heart rate (in a valid range of 0.7Hz - 2.5Hz). Tarassenko \cite{tarassenko2014non3in1} et. al propose a pipelined approach that uses face tracking and pole selection mechanism to estimate heart rate, respiration rate, and oxygen saturation. However, these methods are susceptible to noise, motion, and lighting conditions of the environment.

% \vspace{2mm}

Advancements in deep learning \cite{eccv2020meta,neurips2020multi,icassp_da_2020,nowara2020benefit,Sabokrou2021,metaphys} have made it possible to achieve remarkable performance in the task of video-based human physiological estimation. One of the early popular methods in this direction is DeepPhys \cite{eccv2018deepphys}. This model is trained on many facial videos in a supervised fashion, where ground truth blood volume pulse was used as the label. The key idea is to use separate branches for modeling motion and appearance. The latter branch aids the former by providing attention over facial pixels. Similarly, a more recent method MTTS-CAN \cite{neurips2020multi} used the attention mechanism in conjunction with Temporal Shift Modules \cite{tsm} to achieve real-time performance on the task. In \cite{eccv_20_video}, the authors demonstrated an effective method to disentangle spatio-temporal representation of the video called MSTmaps into noisy signals and physiological signals. In summary, deep learning methods have demonstrated reliable performance on datasets containing relatively stable physiological signals. In the V4V challenge, we introduce a dataset in which the physiological signals vary significantly due to the elicitation of spontaneous emotions.

\section{\textit{V4V dataset} curation}

As part of the V4V challenge, we curated a database called the \textit{V4V dataset} by carefully selecting subjects from the Multimodal Spontaneous Emotion database (BP4D+) \cite{mmse_bp4d} along with a number of new subjects that are collected as BP4D+ extension. In this section, we describe the data collection process, distribution of subjects, and data annotation process used for creating the \textit{V4V dataset}. In this section, we also describe the significance of the dataset for V4V challenge since the dataset contains challenging variations in the physiological signals induced through emotion elicitation.

\begin{figure}[t]
    \centering
    \includegraphics[width=\columnwidth]{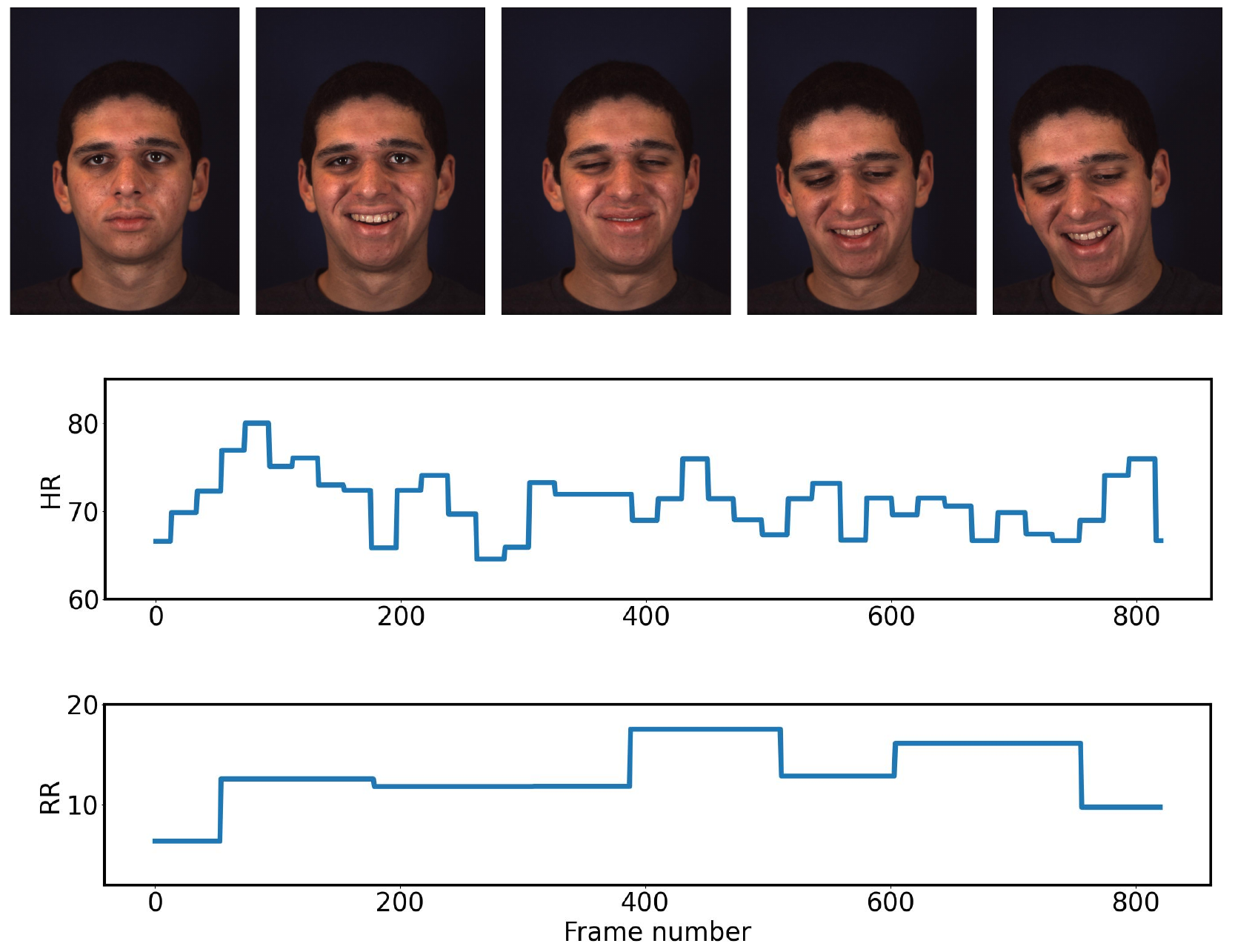}
    \caption{Subject participating in a silly song (T6) task}
    \label{fig:emotion_m53}
\end{figure}

\subsection{Data collection and annotation}

The \textit{V4V dataset} was curated with the goal of obtaining a large-scale emotional corpus for human behavioral and physiological analysis. The dataset is collected at Binghamton University and includes subjects of age groups ranging from 18 to 66. It has subjects from diverse ethnicities/racial ancestries - African American, White, Asian (East and Middle-east), Hispanic/Latino, Native American. There are 179 subjects in total with a maximum of 10 experimental tasks per subject. Each task was specifically designed to induce specific emotions among participants.

\begin{figure}[t]
    \centering
    \includegraphics[width=\columnwidth]{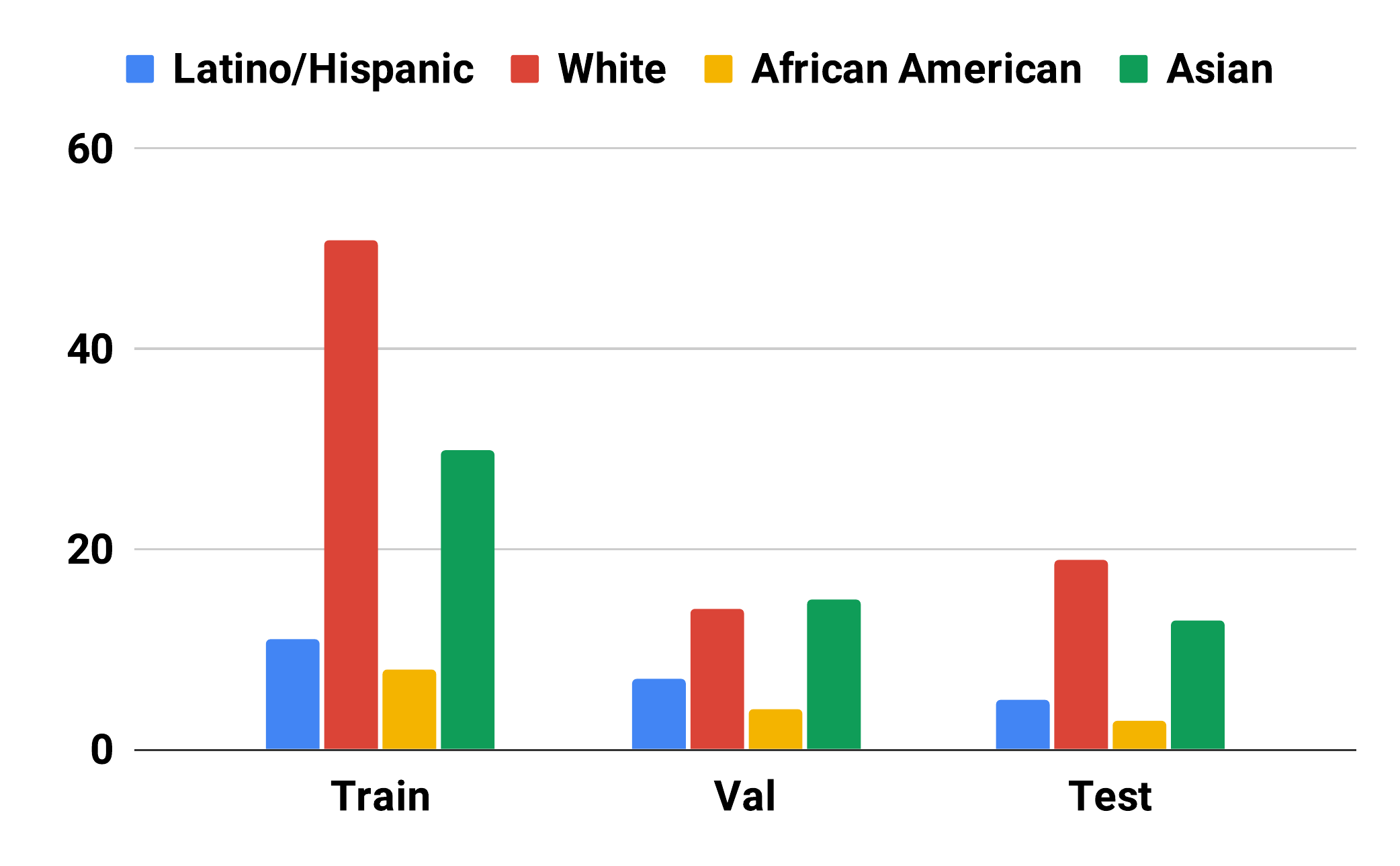}
    \caption{Distribution of videos according to the ethnicity of subject for train, validation, and test subjects.}
    \label{fig:ethnic}
\end{figure}

For recording the facial videos of each subject, a 3D Dynamic imaging system (Di3D) was used. All the videos were recorded with a resolution of $1040\times1392$ pixels and a fixed frame rate of 25 FPS in portrait mode. The Di3D system also has a symmetric lighting system that is used as the light source for capturing the videos. A board has been placed in the background while recording the video to limit any background motion and noise. 

% \vspace{2mm}

For collecting human physiological data, the Biopac MP150 system was used. The specification of the device is as follows:

\begin{itemize}
    
\item Blood pressure: For monitoring the blood pressure, Biopac NIBP100D system with a measurement range of (-25mmHg, 300mmHg) was used. It contains a finger unit and an inflatable cuff that can be placed on the arm to measure blood pressure. The device recorded high-quality measurements of systolic and diastolic blood pressure and also recorded the continuous blood pressure waveform at 1000Hz. 
\item Heart Rate (HR) measurement: We used off-the-shelf software called Biopac AcqKnowledge to derive HR measurement from the continuous blood pressure signal. This was achieved by performing noise removal, followed by peak-to-peak time calculation. %The exact specifics of the HR computing software are unclear since the Biopac AcqKnowledge is proprietary software. 
The software used an HR range preset to (40, 180) beats-per-min. The HR signal is then downsampled and synchronized with each frame.

\item Respiration Rate (RR) measurement: The respiratory signal was captured by using the Biopac Respiration Belt. Similar to the extraction of HR, we used the Biopac Acqknowledge software to extract the respiration rate of the subject by computing peak-to-peak time with RR range preset to (4, 20) breaths-per-min.

\end{itemize}

In order to synchronize the camera and physiological devices, a controller was used to trigger the start of video capture and physiology measurement simultaneously.

\begin{table}[t!]
    \centering
    \setlength\tabcolsep{2.5pt}
    \caption{Ten tasks and the target emotion}
    \begin{tabular}{|c|c|c|}
    \hline
    Task & Activity performed & Emotion induced \\
    \hline
      T1 & Funny joke & Happiness \\
      T2 & Watch 3D avatar of self & Surprise \\
      T3 & 911 emergency call & Sadness \\
      T4 & Sound & Surprise \\
      T5 & True / False question & Skepticism \\
      T6 & Silly song & Embarrassment \\
      T7 & Dart threat & Fear \\
      T8 & Cold presser & Pain \\
      T9 & Complaining against participant & Angry \\
      T10 & Odor experience & Disgust \\
      \hline
      
    \end{tabular}
    
    \label{tab:t10}
\end{table}

\begin{table}[t!]
    \centering
    \caption{\textit{V4V Dataset}: Data split}
    \begin{tabular}{|c|c|c|c|}
        \hline
        Data Fold & \makecell{Number of \\subjects} & \makecell{Number of \\Videos} & \makecell{Average video \\ length (in sec)} \\
        \hline
        Train & 100 & 724 & 44.2 \\
        \hline
        Validation & 39 & 276 & 42.9 \\
        \hline
        Test & 40 & 358 & 45.8 \\
        \hline
         
    \end{tabular}
    
    \label{tab:splits}
\end{table}

\subsection{Emotion elicitation protocol of the \textbf{\textit{V4V dataset}}}

Each subject participated in 10 different tasks that were carefully designed to evoke a specific emotion. It is known that an increased emotional activity often alters the human physiological signals \cite{heartrateemotion2019}. For example, fear arousal spontaneously increases the heart rate and respiration rate of a subject. To achieve this goal, a skilled interviewer was hired to conduct the tasks. %We now describe the data collection protocol for emotional elicitation.

% \vspace{2mm}

As shown in Table \ref{tab:t10}, the tasks included interpersonal communication, film watching, cold pressure, and physical activities. There is also a smooth transition in the emotions induced across the 10 tasks.  There was a brief pause between any two tasks for self-reporting purposes.

% \vspace{2mm}

First, the interviewer ensured that the participant felt comfortable and relaxed at the start of data collection by advancing a joke (T1). Then the subject was shown their own 3D avatar to invoke the feeling of surprise (T2). Next, the T3 task required participant to watch a short film of a 911 emergency call to elicit a feeling of sadness. In the T4 task, a loud noise was played to startle and surprise the participant. The interviewer then induced skepticism by advancing a question (T5), followed by arousal of embarrassment when the subject was required to conduct a silly task (T6). In the T7 task, the interviewer invoked fear in the participant by threatening to throw a dart at the subject. In T8, the participant was required to submerge their hand into the ice water which invoked physical pain. The interviewer pretended that the participant demonstrated poor performance in task T8 and complained to the participant to evoke anger (T9). In the final task, T10, the subject experienced a smelly odor to evoke the feeling of disgust. At the end of each task, the subject was asked to report the emotions experienced from a list of choices and also rate the emotional intensity in a 5-point rating style. 

% \vspace{2mm}

Owing to the carefully designed emotion elicitation, the \textit{V4V dataset} contains challenging intra-video physiological variations. Further, due to the nature of the tasks T6-T10, they often are associated with large head movements ($>10$ deg) adding an additional element of challenge for physiological estimation. Therefore, \textit{V4V dataset} offers desirable elements for benchmarking approaches effectively.

\subsection{Post-processing}

After creating the ground truth, we eyeballed each of the heart rate and respiration rate sequences to discard any video that had noisy readings, e.g. Shaking the contact-based device during emotion elicitation tasks. We ensured that every video had exactly 1 HR and 1 RR reading per frame of the video after aligning the physiological signals. After processing the data as described, we obtained 179 subjects and 1358 videos with heart rate and respiration rate readings. 

\begin{figure}[t]
    \centering
    \includegraphics[width=\columnwidth]{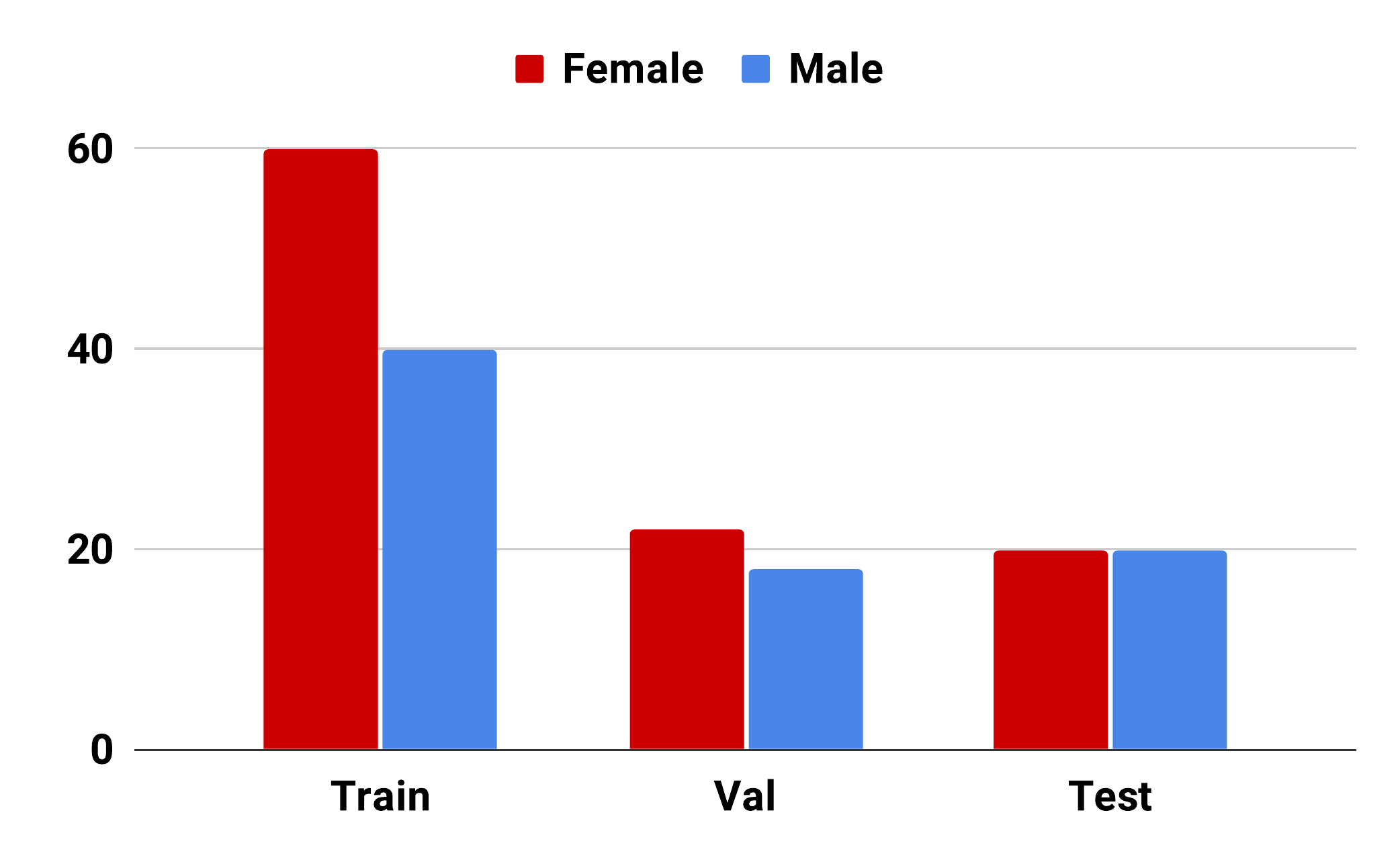}
    \caption{Distribution of videos according to subject gender
    for train, validation, and test splits.}
    \label{fig:gender}
\end{figure}

\subsection{V4V Challenge phases and dataset split}

The 1st V4V challenge was organized in two phases. In the first phase, the participants used the validation set to improve performance of their methods, and in the second phase, the participants evaluated their method on the test set. A public leaderboard was maintained in both the phases indicating scores obtained by the methods on the evaluation metrics described in Sec. \ref{sec:eval}.

As shown in Table \ref{tab:splits}, we used 1000 videos in phase 1 of the challenge where the training set included 724 videos and validation set included 276 videos. At the start of phase-1, we released the training dataset and validation set videos. The training set contains the compressed videos, heart rate signal, respiration rate signal, and raw 1000 Hz blood pressure waveform. At the start of phase-2, we released the test set videos (a total of 358 videos) and validation-set ground truth as well. 

In order to ensure that the methods performed fairly across different population groups, we used gender and ethnicity information self-reported by the subjects to create the data folds. %{\color{blue}\textbf{TODO} First we grouped all participants by group and gender}
Fig. \ref{fig:ethnic} shows the ethnic distribution of subjects of train, validation, and test set. We tried to balance the ethnic distribution keeping them similar across the data splits to avoid the biases brought by skin tone and affective attributes.
Fig. \ref{fig:gender} presents the distribution of videos according to the gender of the subject, which has been adjusted to have near-balanced distribution, with the number of female subjects slightly more than the number of male subjects in the train and validation set.

\section{Evaluation metrics}
\label{sec:eval}

In the existing literature, Mean Absolute Error (MAE), Root Mean Squared Error (RMSE), and Pearson Correlation Coefficient (R) have been used commonly for evaluating the efficacy of the proposed method. The caveat is that almost all of the existing works construct 30-second non-overlapping segments and predict a single HR/RR value per segment. While this is a good measure of the performance of the method, it has some potential drawbacks.

% \vspace{2mm}

Unlike other physiological datasets, the \textit{V4V dataset} consists of short video clips that have significant intra-video variations in the HR and RR (as seen in Fig. \ref{fig:emotion_f74} and Fig. \ref{fig:emotion_m53}) for the duration of the video. Therefore, an accurate method should be able to predict HR and RR at a more fine-grained frame-level rather than at a segment-level. Further, contact based devices predict the heart rate continuously, instead of one prediction for every 30s. This is especially useful in diagnosing Atrial fibrillation (Afib) \cite{npjafib2020} which is associated with high Heart Rate Variability. Based on these observations, we propose a metric that computes error by taking into account the frame-level HR/RR.

% \vspace{2mm}

In the 1st V4V challenge, we employ the following three metrics - MAE, RMSE, and R - at a frame-level rather than at segment-level. We denote these granular evaluation metrics as cMAE, cRMSE, and cR respectively.

\begin{equation}
    cMAE=\frac{\Sigma_i|\widehat{HR}_i -HR_i|}{N}
\end{equation}

\begin{equation}
    cRMSE=\sqrt{\frac{(\Sigma_i|\widehat{HR}_i -HR_i|^2)}{N}}
\end{equation}

\begin{equation}
    cR = \mathbf{PearsonCorrelation}(\widehat{HR}, HR)
\end{equation}

where $HR_i$ is the ground truth heart rate of the frame $i$ in the test set, $\widehat{HR}_i$ is the predicted heart rate for the frame $i$ in the test set and $N$ is the total number of frames in the test set. Similarly, we define equivalent evaluation metrics for Respiration Rate using $RR_i$ and $\widehat{RR}_i$.

\section{Methods used in the V4V challenge}

\begin{table}[t!]
    \centering
    
    \caption{Results obtained on the \textit{V4V dataset} (test set) sorted by cMAE (lower is better $\downarrow$)}
    \setlength{\tabcolsep}{3pt}

    \begin{tabular}{|l|c|c|c|}
        \hline
        Approach & cMAE ($\downarrow$) & cRMSE ($\downarrow$) & cR ($\uparrow$)\\
        \hline
        Stent et. al \cite{shisen2021selfsup} & 9.22 & 14.18 & 0.47 \\
        \hline
        Hill et. al \cite{brian2021hybrid} & 9.37 & 14.59 & 0.44 \\
        \hline
        Kossack et. al \cite{kossack2021regionbased} & 10.15 & 15.38 & 0.44 \\
        \hline
        Ouzar et. al \cite{ouzar2021lcoms} & 11.60 & 14.90 & 0.20 \\
        \hline
        \hline
        Baseline (Green \cite{green2008remote}) & 15.45 & 20.73 & 0.05 \\ 
        \hline 
         
    \end{tabular}
    
    \label{tab:v4vresults}
\end{table}

There were four teams that participated in the V4V challenge. In Table. \ref{tab:v4vresults} results obtained by different teams and a baseline (Green) method  \cite{green2008remote,mcduff2019iphys} have been listed.

\subsection{Estimating Heart Rate from Unlabelled Video}  Stent et. al \cite{shisen2021selfsup} use a self-supervised approach based on \cite{iccv2021selfsup} to overcome challenges that are typically faced by a supervised approach such as imprecise and noisy data collection owing to complex ground truth capturing setup, proprietary hardware computations, etc. Further, by virtue of the self-supervised nature of the proposed method, it can also tackle any domain shift problem that arises during the test phase. First, an image of size 192$\times$128 is extracted by using a face detector \cite{facedetZhang2017s3fd} and padded with an additional 25\% buffer. Then a PPG estimator consisting of 3DCNN \cite{3DCNNyu2019remote} is used to extract the rPPG signal. 

The core idea of this approach is to formulate a self-supervised framework by utilizing a video resampling module to augment the dataset with new heart rate labels. E.g., the heart rate of the subject can be increased by squeezing video. Similarly, stretching the video decreases the heart rate. The module is not only used to increase/decrease the heart rate by controlling the speed for video, but it is also used to resample the output rPPG signal with corresponding inverse frequency. Based on this idea, positive and negative pairs are constructed for an effective contrastive learning \cite{mocoHe2020momentum,simclr2020simple}.

Owing to the self-supervised nature of the proposed method, the method is able to train on the test samples directly. The paper also implements several hand-crafted tricks to further improve performance of the their method. The authors employ specialist models to account for multi-mode lighting conditions in the test set and also propose a confidence model to mitigate errors caused by faulty PPG prediction.

\subsection{Beat-to-Beat Cardiac Pulse Rate Measurement From Video} Hill et al. \cite{brian2021hybrid}  proposed a hybrid-CAN-RNN framework by incorporating a bi-directional GRU on the top of CAN \cite{neurips2020multi}. The original Hybrid-CAN has two branches: (1) the appearance branch, and (2) the motion branch. The appearance branch deploys 2D convolutions to extract the spatial skin features. The temporal module leverages 3D convolutions to capture the temporal relationships. An attention module is used to bridge the connection between motion and appearance branch, and the generated attention masks make the network focus on the useful signals.

In addition, two layers of GRU (the first is bi-directional) are used for learning the longer-term rPPG waveform transitions, and the reason is that the convolutional layers can only capture local spatial and temporal features. Besides, a synthesized dataset is used to improve the model generalization ability, and other datasets such as AFRL \cite{estepp2014recovering} and UBFC \cite{bobbia2019unsupervised} are used for training. The variety of the training data improves the model generalization ability, which helps to deal with more complicated environmental biases.

\subsection{LCOMS Lab’s approach to the Vision For Vitals (V4V) Challenge} To model the spatial and temporal features concurrently, Ouzar et al. \cite{ouzar2021lcoms} make use of the residual connected 3D Depthwise Separable Convolution layers. The depth-wise separable convolution significantly reduces the parameters and saves the computational time, and achieves satisfactory results. Moreover, their method can run in real-time both in CPUs and GPUs. 

Different from the others that directly feed the images with the background to the deep networks, this paper uses a precise face swapping-based segmentation method \cite{nirkin2018face} to exclude the background and only keep the facial regions. The paper shows that most of the heart rate values are inside the range of 70 BPM to 90 BPM. And to tackle this data imbalance issue, they performed an offline data augmentation to the sequences that have minority heart rate values. 

Although the framework is end-to-end trainable and is superior in speed and simplicity, the basic idea is still treating each task as a one-stage regression problem. The model predicts one average heart rate value in two seconds video segment. Proposing a solution to predict frame-wise heart rate (e.g. combining frame-specific features with segment features) can be a future work of this paper.

\subsection{Automatic region-based heart rate measurement using remote photoplethysmography} Kossack et. al \cite{kossack2021regionbased} utilize a popular classical method called ``plane orthogonal to skin" (POS) \cite{wang2016algorithmic} to measure the heart rate. The core idea of this V4V submission is to extract a signal from the region of the face that has the strongest signature of rPPG in it. In order to achieve this goal, the face region is divided into five subregions - forehead, right cheek, left cheek, nose, and the entire face is considered as the fifth subregion. For each of these ROIs, a score is computed and the ROI with the best score is used to determine the heart rate of the subject.

First, POS projection is applied to each subregion to extract the rPPG signal followed by FFT and bandpass filtering to obtain a power spectrum density for each subregion. Next, a scoring function is designed for the determination of the best region of interest. This function is dependant on two parameters, (1) Maximum magnitude ($Mmax$) and (2) Row wise sum of the correlation matrix of 5 rPPG signals ($Csum$). The sum of $Mmax$ and $Csum$ yields the final score which is used to determine the best region of interest. Since this method does not take a deep learning approach, the method can be used directly on the test set.

\subsection{Results}  Various approaches are evaluated for cMAE, cRMSE, cR as shown in the Table \ref{tab:v4vresults}. It is interesting to note that the self-supervised formulation used by \cite{shisen2021selfsup} is on top of the leaderboard demonstrating good performance on continuous evaluation metrics. However, their approach also involves carefully handcrafted tricks. 
The method presented in \cite{brian2021hybrid} used additional datasets for training and obtained cMAE of $9.37$. All participants chose to participate only in the HR sub-challenge. 
In summary, the results obtained by different methods indicate that continuous prediction is still a challenging task and there is scope for further improvements.

\section{Conclusion}
In this paper, we have presented the first Vision-for-Vitals challenge for benchmarking the performance of different rPPG methods on a newly curated large-scale dataset called \textit{V4V dataset}. The dataset contains desirable attributes necessary for benchmarking various approaches including challenging elements such as spontaneous behavior and varied HR/RR signals. Further, in order to benchmark different methods effectively, we evaluate various approaches using granular frame-level error metrics rather than segment-level error metrics employed by previous methods. The results show that there is room for further improvement in methods and evaluation protocols used for non-contact video-based human physiological estimation. 

Recent methods introduced in the area of self-supervised learning \cite{mocoHe2020momentum,simclr2020simple}, semi-supervised learning \cite{kundu2020unsupervised,revanur2021semi}, domain adaptation \cite{kundu2020towards,cida_eccv,venkat2021classifier} and transformer architectures \cite{vaswani2017attention,dosovitskiy2020image} have pushed state-of-the-art across a broad range of computer vision tasks. These techniques could also improve the performance of human physiology estimation.

\section*{Acknowledgement}

\noindent This research was supported in part by the Bill \& Melinda Gates Foundation (BMGF) under grant INV-002154, the  NSF under grant CNS-1629898, and the Center of Imaging, Acoustics, and Perception Science (CIAPS) of the Research Foundation of Binghamton University.

{\small
\bibliographystyle{ieee_fullname}
\bibliography{egbib}
}

\end{document}